\documentclass[prb,twocolumn,amsmath,amssymb,superscriptaddress,letterpaper]{revtex4}

\usepackage{graphicx}           % Extended Graphics package.
\usepackage{times}              % Font is Times New Roman. Others are also available

\usepackage[pdftex]{hyperref} % REMOVE WHEN COMPILING FOR REAL WITH CONDMAT

\newcommand{\ket}[1]{|#1\rangle}

\begin{document}

\title{Scalable in-situ qubit calibration during repetitive error detection}

\author{J. Kelly}
\affiliation{Google Inc., Santa Barbara, California 93117, USA}

\author{R. Barends}
\affiliation{Google Inc., Santa Barbara, California 93117, USA}

\author{A. G. Fowler}
\affiliation{Google Inc., Santa Barbara, California 93117, USA}

\author{A. Megrant}
\affiliation{Google Inc., Santa Barbara, California 93117, USA}

\author{E. Jeffrey}
\affiliation{Google Inc., Santa Barbara, California 93117, USA}

\author{T. C. White}
\affiliation{Google Inc., Santa Barbara, California 93117, USA}

\author{D. Sank}
\affiliation{Google Inc., Santa Barbara, California 93117, USA}

\author{J. Y. Mutus}
\affiliation{Google Inc., Santa Barbara, California 93117, USA}

\author{B. Campbell}
\affiliation{Department of Physics, University of California, Santa
Barbara, CA 93106, USA}

\author{Yu Chen}
\affiliation{Google Inc., Santa Barbara, California 93117, USA}

\author{Z. Chen}
\affiliation{Department of Physics, University of California, Santa
Barbara, CA 93106, USA}

\author{B. Chiaro}
\affiliation{Department of Physics, University of California, Santa
Barbara, CA 93106, USA}

\author{A. Dunsworth}
\affiliation{Department of Physics, University of California, Santa
Barbara, CA 93106, USA}

\author{E. Lucero}
\affiliation{Google Inc., Santa Barbara, California 93117, USA}

\author{M. Neeley}
\affiliation{Google Inc., Santa Barbara, California 93117, USA}

\author{C. Neill}
\affiliation{Department of Physics, University of California, Santa
Barbara, CA 93106, USA}

\author{P. J. J. O'Malley}
\affiliation{Department of Physics, University of California, Santa
Barbara, CA 93106, USA}

\author{C. Quintana}
\affiliation{Department of Physics, University of California, Santa
Barbara, CA 93106, USA}

\author{P. Roushan}
\affiliation{Google Inc., Santa Barbara, California 93117, USA}

\author{A. Vainsencher}
\affiliation{Department of Physics, University of California, Santa
Barbara, CA 93106, USA}

\author{J. Wenner}
\affiliation{Department of Physics, University of California, Santa
Barbara, CA 93106, USA}

\author{John M. Martinis}
\affiliation{Google Inc., Santa Barbara, California 93117, USA}
\affiliation{Department of Physics, University of California, Santa
Barbara, CA 93106, USA}

\begin{abstract}
We present a method to optimize qubit control parameters during error detection which is compatible with large-scale qubit arrays. We demonstrate our method to optimize single or two-qubit gates in parallel on a nine-qubit system.  Additionally, we show how parameter drift can be compensated for during computation by inserting a frequency drift and using our method to remove it. We remove both drift on a single qubit and independent drifts on all qubits simultaneously. We believe this method will be useful in keeping error rates low on all physical qubits throughout the course of a computation. Our method is $O(1)$ scalable to systems of arbitrary size, providing a path towards controlling the large numbers of qubits needed for a fault-tolerant quantum computer.
\end{abstract}

\maketitle

A fault-tolerant quantum computer protects a quantum state from the environment through the careful manipulations of millions of physical qubits~\cite{fowler2012surface}. In such a computer, each qubit must reliably perform a series of quantum logic gates~\cite{chow2012universal, harty2014high, barends2014superconducting, ballance2015laser} to detect and negate errors~\cite{kelly2015state, corcoles2015demonstration, chow2014implementing}. However, operating such quantities of qubits at the necessary level of precision is an open challenge, as optimal control parameters can vary between qubits~\cite{gambetta2015building} and drift in time~\cite{bialczak20071}. Here we present a method to optimize control parameters and counteract system drift that scales to arbitrary numbers of qubits, that can be performed during computation with no additional overhead in time. The presented approach  is in principle applicable to any code that repetitively detects errors using small groups of qubits. We implement our method on a superconducting nine-qubit device performing repetitive error detection demonstrating how parameters for single and two-qubit gates can be scalably optimized in parallel. Additionally, we show how independent parameter drifts on each qubit can be tracked and removed during computation. These results provide a path forward to controlling the large-scale qubit arrays needed for fault-tolerant quantum computation. 

Finding and maintaining optimal control parameters of a continuously running quantum computer is of great interest as useful algorithms on future computers will likely require large arrays of qubits operating without fail for days at a time. An ideal optimal quantum control method would run in parallel throughout the computation, to track and compensate for the unavoidable drifts in the system~\cite{miquel1997quantum}. Optimal quantum control~\cite{shapiro2003principles} has a rich history in state transfer~\cite{khaneja2005optimal, bardeen1997feedback}, creating macroscopic quantum states~\cite{koch2004stabilization, hohenester2007optimal}, optimizing quantum gates~\cite{sporl2007optimal, kelly2014optimal, egger2014adaptive, egger2014optimized}, and controlling many-qubit systems~\cite{zhang2011coherent, zhang2011experimental}. Conventional methods such as tomography~\cite{o2004quantum} and randomized benchmarking \cite{magesan2011scalable} use the final state of the system as a metric to evaluate the performance of a control sequence~\cite{calarco2004quantum}, and would require interruption of the necessary error detection algorithm. Thus, these methods do not extend to a continuously running quantum computer, as control is done on-the-fly, the qubits which store the quantum data (``data qubits'') may not be measured, and errors introduced while exploring parameter space may lead to logical failure. Perhaps most importantly, real-time device performance can only be assessed in detection events, the outputs of error detection operators.

We present a method that uses detection events -- the rate that errors occur -- as a metric. Using codes where error detection and propagation is bounded, we can partition a system into qubit groupings that can be tuned independently. We feedback the error rate from a grouping to improve control parameters for gates contained within that grouping. By choosing finite patterns of groupings, we can independently optimize every control parameter of every qubit and retain $O(1)$ scaling with system size. Since we use error detection to inform our control parameters, we are guaranteed optimal performance, and there is no need to interrupt error detection to perform calibration. Additionally, the qubits do not need to be operating below threshold to use this technique. We call our method Active Detection Event Parameter Tuning (ADEPT).

\begin{figure}[!]
    \centering
    \includegraphics[width=0.48\textwidth]{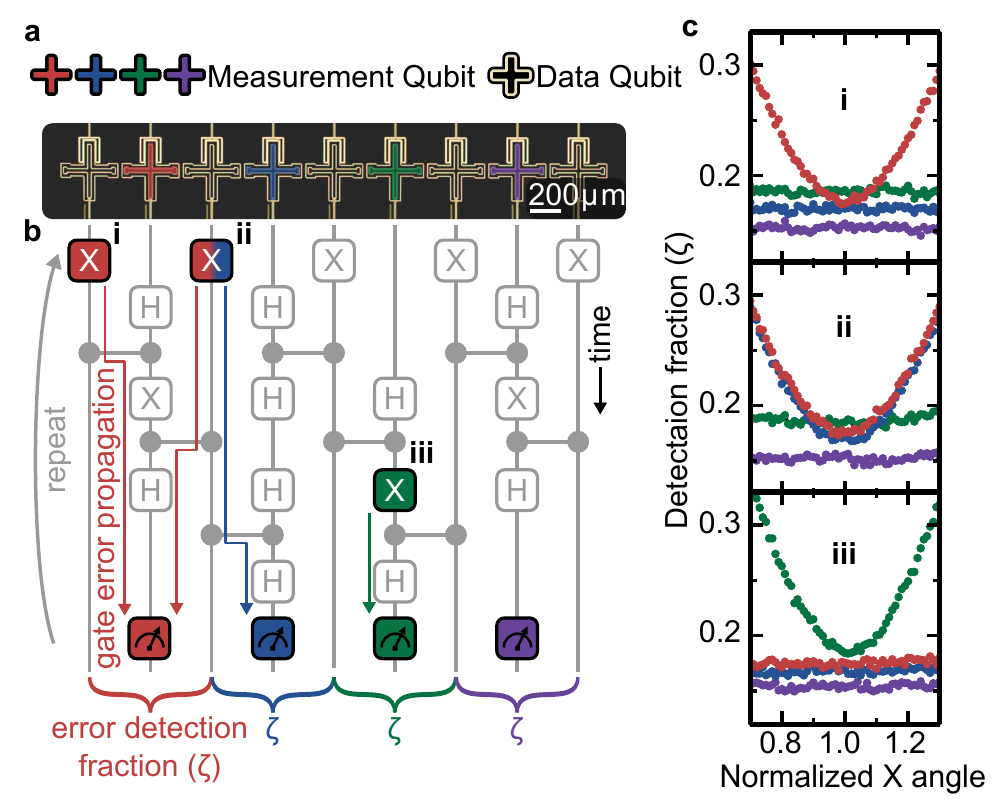}
    \caption{(Color) \textbf{Error detection circuit and gate error detection.} (a) Linear chain of nine superconducting qubits with nearest neighbor coupling. (b) Error detection circuit for the bit-flip repetition code, where data qubits (black and gold) hold the quantum state, and measurement qubits (colored) are used to detect errors. Here we perform three experiments \textbf{i}-\textbf{iii}, where rotation angles are tuned from their optimal values and errors are recorded. Gate-induced bit errors on data qubits (\textbf{i} and \textbf{ii}) are copied to measurement qubits through CZ gates where they are detected as detection events, see text. Gate errors on measurement qubits (\textbf{iii}) are localized to those qubits. Hadamard gates are physically implemented with variable phase $\frac{\pi}{2}$ rotations. (c) Detection event fraction $\zeta$ $vs$ variable angle X gates on data (\textbf{i} and \textbf{ii}) and measurement (\textbf{iii}) qubits. Each data point is the average of 6,000 instances of eight rounds of detection. Change in gate parameters from their ideal increases $\zeta$. Change in $\zeta$ is localized to measurement qubits near the gate being varied, as $\zeta$ for unaffected qubits is constant.}
    \label{fig:circ}
\end{figure}

We demonstrate ADEPT on a nine-qubit superconducting quantum processor.  It consists of a thin superconducting film of Aluminum on sapphire, which is lithographically defined into a linear chain of Xmon transmon qubits~\cite{koch2007charge, barends2013coherent} with individual control and readout (Fig~\ref{fig:circ}a). Single qubit rotations are performed with microwave pulses at the qubit frequency (4-6 GHz), and a current bias can be applied to bring neighboring qubits into resonance, enabling a controlled-Z (CZ) gate~\cite{barends2014superconducting}. Measurement is achieved using dispersive readout~\cite{schuster2005ac, wallraff2005approaching, jeffrey2014fast, mutus2014strong}.

For our error detection algorithm, we choose to work with the repetition code. The nearest neighbor coupling makes it a natural choice for our architecture, it has been experimentally demonstrated to operate below the threshold for error correction, and is the one-dimensional primitive of the two-dimensional surface code~\cite{kelly2015state}. The repetition code detects bit-flip errors: here, entangling gates are used to copy bit errors on data qubits, which store the quantum state, onto neighboring ancilla measurement qubits (Fig~\ref{fig:circ}b \textbf{i}, \textbf{ii}), where they can be detected. Bit errors on measurement qubits will change their state, but will not propagate back to data qubits (Fig~\ref{fig:circ}b, \textbf{iii}). Errors are detected by repeatedly performing the error detection circuit, a $\hat{Z}\hat{Z}$ stabilizer, and analyzing the measured states of the measurement qubits (Fig~\ref{fig:circ}c). In a future quantum computer, these operations would be running continuously without interruption. While this is technically challenging today, we emulate the performance of $N$ rounds of error detection by initializing the system into the logical $\ket{0}$ state (which has similar performance to the logical $\ket{1}$, see ref.~\onlinecite{kelly2015state}), run eight rounds of error detection, and end the code. Then, we repeat this process to gather statistics and accumulate $N$ total rounds of detection~\cite{supp}.

We now discuss how we process the measured qubit states into error detection events. In the presence of no error, measurement qubits will report a string of repeated or alternating states, depending on the states of the neighboring data qubits. In the presence of an error on the measurement qubit or a neighboring data qubit, the pattern of states will flip between the repeated or alternating pattern. This is known as a detection event, and indicates the presence of a nearby error (see ref~\onlinecite{kelly2015state} for more detail). The error rate of the system is thus directly related to the detection event fraction $\zeta$, the fraction of measurements that are detection events. The $\zeta$ presented are consistent with the below threshold behavior demonstrated in ref~\onlinecite{kelly2015state}. 

As an experimental demonstration of the relation between gate errors and detection events, we have inserted error by tuning rotation angles away from the optimum for specific qubits. The results are shown in Fig~\ref{fig:circ}d. The errors from a miscalibrated gate on data qubits (Fig~\ref{fig:circ}d \textbf{i} and \textbf{ii}) are copied onto their neighboring measurement qubits (red, and red/blue respectively). Errors from miscalibrated gates on a measurement qubit are localized to that qubit (Fig~\ref{fig:circ}d \textbf{iii}).

Note that there is a clear connection between the parameters of a gate and $\zeta$ of nearby measurement qubits. We see that away from the optimal rotation angle for an X gate (rotation around X-axis in the Bloch sphere representation), $\zeta$ increases, making it a natural metric to improve the gate parameters. Second, we see that the change in $\zeta$ is local: if the gate parameters are adjusted on a data qubit (i and ii), only the neighboring measurement qubits change $\zeta$. If the gate parameters on a measurement qubit are adjusted (iii), only that measurement qubit has a change in $\zeta$. In both cases, unaffected qubits see no change. 

Crucially, the direct, correspondence between detection events and gate errors on nearby qubits implies that we can tune gate parameters in parallel whenever the measurement qubits that pick up these gate errors do not overlap. 

\begin{figure}[!]
    \centering
    \includegraphics[width=0.48\textwidth]{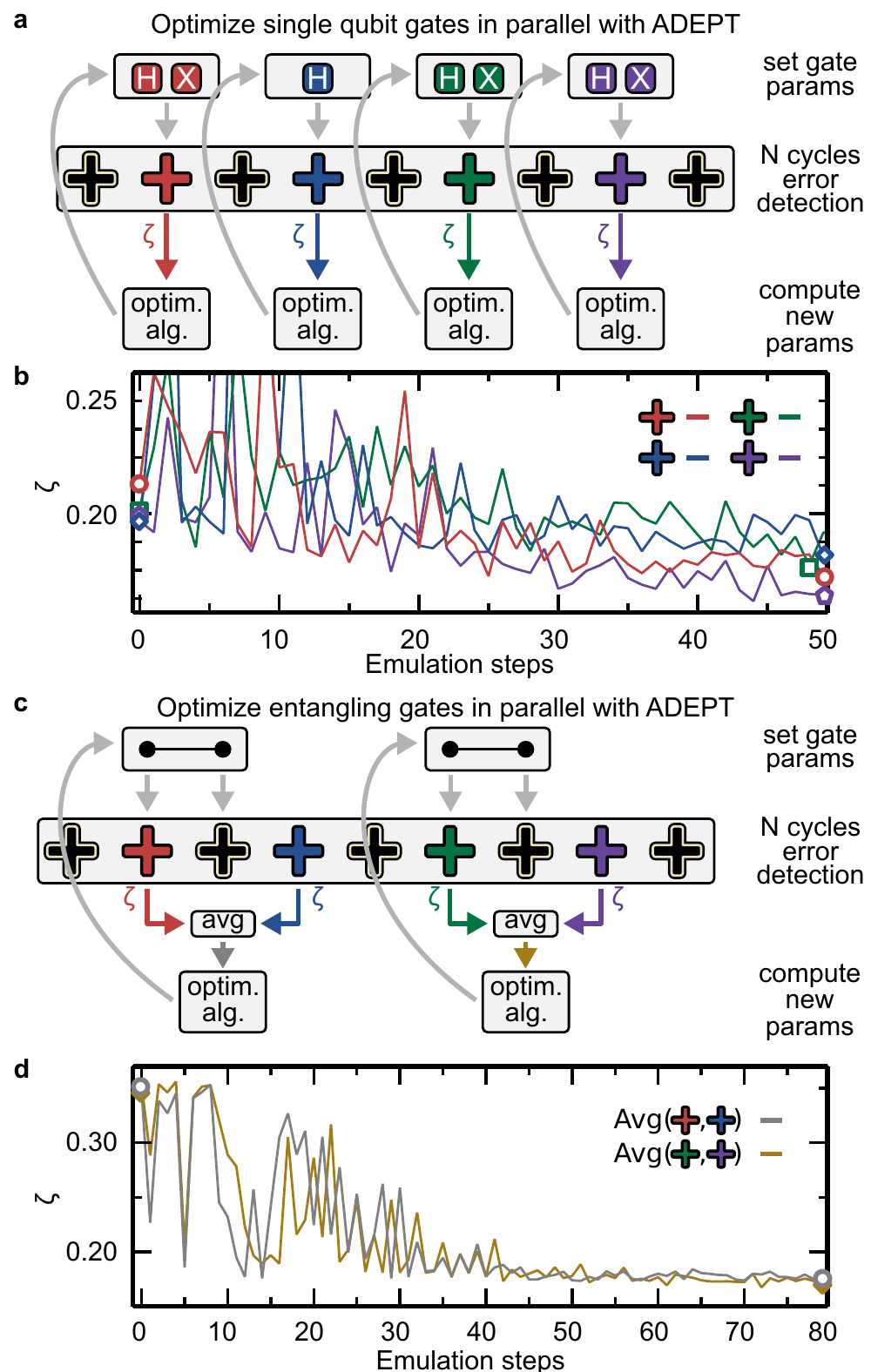}
    \caption{(Color) \textbf{Parallel gate optimization using independent hardware groupings.} (a) Schematic of ADEPT for optimizing gates on measurement qubits. Gate parameters are chosen and error detection is run to determine $\zeta$ that is used as an error metric for the Nelder-Mead optimization algorithm, which chooses new gate parameters. (b) Parallel optimization of measurement qubit gate parameters. Each measurement qubit is optimized independently, and each iteration is 4,500 instances of eight rounds of detection. Average $\zeta$ is improved from 0.202 to 0.179. (c) Using ADEPT to optimize CZ gates. As CZ gates include data qubits, both neighboring measurement qubits will change $\zeta$ with changing gate parameters. To avoid error overlap, we optimize CZ that are well separated (see text). (d) Parallel Nelder-Mead optimization of two CZ gates. The metric is the average $\zeta$ of the measurement qubits that neighbor the data qubit in the CZ. Each iteration is 4,500 instances of eight rounds of detection. Average $\zeta$ is improved from 0.350 to 0.172.}
    \label{fig:gates}
\end{figure}

In Figure~\ref{fig:gates}, we demonstrate how $\zeta$ can be fed back to improve gate parameters in parallel. As a first experiment we optimize single qubit gates on all measurement qubits simultaneously (Figure~\ref{fig:gates}a). We use the $\zeta$ of each measurement qubit -- calculated over $N=36,000$ rounds of detection and constituting one ``emulation step'' -- as an error metric for the Nelder-Mead optimization algorithm, and allow the algorithm to adjust the gate parameters such as amplitude and drive frequency independently on each measurement qubit (Figure~\ref{fig:gates}b). We run this for all four measurement qubits simultaneously. After 50 emulation steps, we find that the $\zeta$ of each measurement qubit has been reduced, from an average of 0.202 to an average of 0.179. 

In Figure~\ref{fig:gates}c, we use the same technique to optimize CZ gate parameters performed between measurement and data qubits. In this case, adjusting gate parameters will change the error rate on both data and measurement qubits involved in the CZ. The additional errors on the data qubit will be copied to both of its neighboring measurement qubits through the error detection circuit (Fig.~\ref{fig:circ} \textbf{ii}). Thus, adjusting CZ gate parameters will alter $\zeta$ of two measurement qubits. To ensure that our chosen error metrics do not overlap, we choose to optimize CZ gates that only involve every other measurement qubit, and take the average $\zeta$ of pairs of measurements qubits as the error metric for Nelder-Mead. After 80 emulations steps, we find a decrease in $\zeta$ from 0.350 to 0.172 (Fig.~\ref{fig:gates}).

These data show that we can optimize gates in parallel without ever interrupting error detection. By extending these techniques we can experimentally determine (or theoretically derive for an ideal system~\cite{supp}) a finite set of experiments that can be run to optimize every gate on every qubit while error detection is running~\cite{supp}. As there are a finite number of experiments, ADEPT is fully parallelizable and has $O(1)$ scaling. We would like to point out that optimal qubit parameters will likely vary between physical qubits due to manufacturing variation, and that qubit parameters therefore need to be individually tuned. Our method will be able to perform this task, and scale to the arbitrary numbers of qubits in future processors running error detection. 

\begin{figure}%[!]
    \centering
    \includegraphics[width=0.48\textwidth]{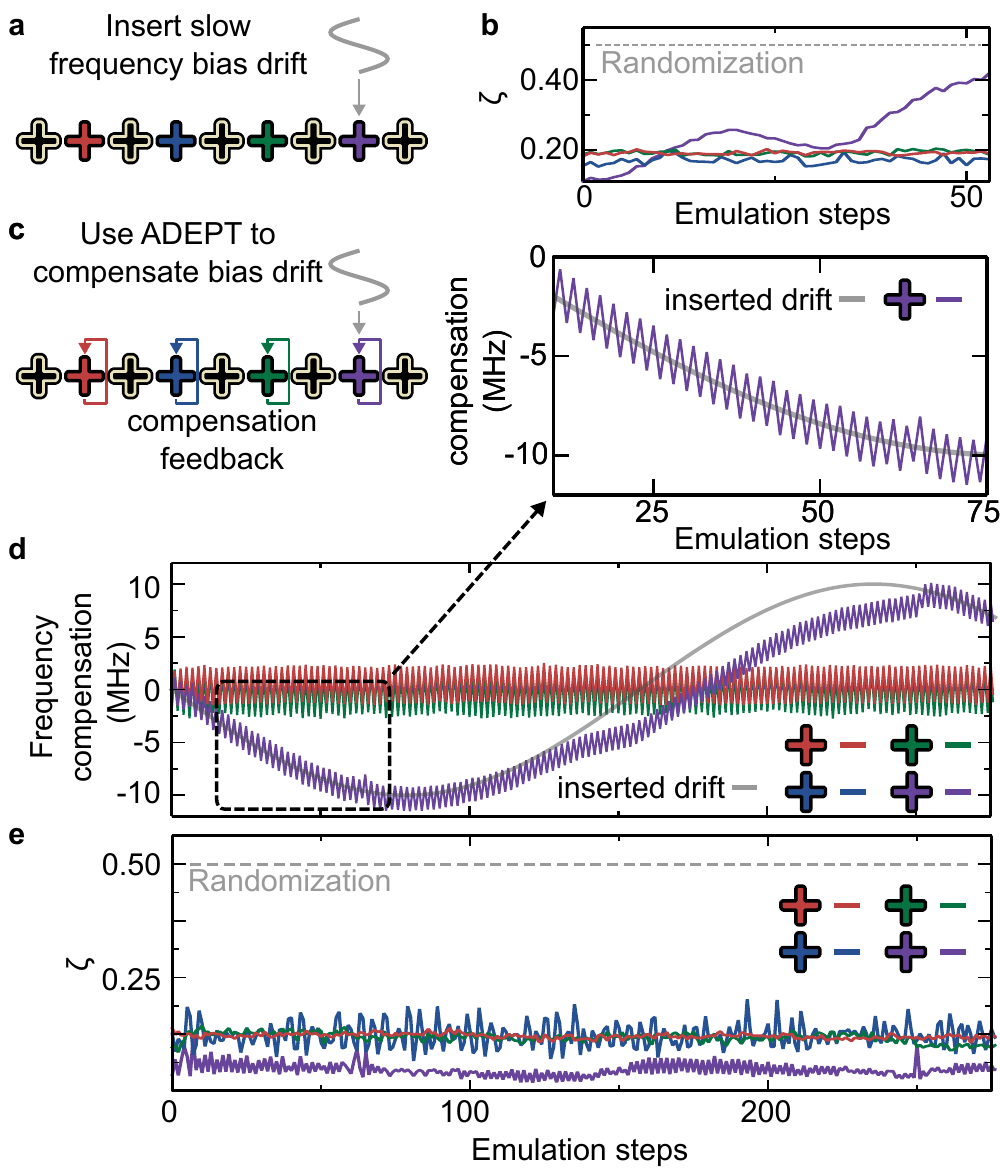}
    \caption{(Color) \textbf{Tracking and negating frequency drift with ADEPT.} (a) A slow sinusoidally varying frequency drift of up to 10~MHz is inserted onto the purple measure qubit. (b) Uncorrected, this will cause $\zeta$ of the purple qubit to double within 20 iterations (with each iteration 12000 instances of eight rounds of detection), and eventually saturate near the randomization limit of 0.5 (c) The frequency drift can be compensated for by feeding $\zeta$ into a following algorithm. The algorithm samples points above and below the ideal bias value, and fits these points to a parabolic error model~\cite{supp}. (d) The drift following algorithm tracks and compensates for the inserted frequency drift through one oscillation. The other algorithms produce no compensation, as no drift is inserted. (e) The measured $\zeta$ remain flat throughout the experiment, in contrast to (b).}
    \label{fig:onebias}
\end{figure}

We have shown ADEPT is adept at finding optimal control parameters, however in real physical systems these ideal parameters can change over time. Given that a future quantum computer will likely perform computations over hours or days~\cite{fowler2012surface}, it is important that parameters remain optimal on this timescale. 

% Julian/Qubit/9Xmon/150823/ [346, 347]
% 6000 stats, 8 cycles, 2 averages
% cycle time is 878 ns

In Figure~\ref{fig:onebias}, we show that ADEPT can be used to compensate for parameter drift. To emulate uncontrolled parameter drift, we insert a slowly varying voltage to the frequency bias of the purple qubit (Figure~\ref{fig:onebias}a), which will induce a large shift on the qubit frequency of up to $\pm$ 10 MHz and updates once each emulation step. We find that after just a 4~MHz shift over 20~rounds of emulation (equivalent to 84.2~ms of non-emulated operation given a 878~ns cycle time~\cite{supp}), $\zeta$ for the purple qubit has more than doubled from 0.11 to 0.26. $\zeta$ eventually exceeds at 0.4 for a 8~MHz frequency error, indicating a near randomization of the measurement qubit output and failure to reliably detect errors. In Fig~\ref{fig:onebias}c, we now feedback $\zeta$ of each measurement qubit to a tracking algorithm that can adjust the frequency bias of that qubit by fitting $\zeta$ to a parabolic error model~\cite{supp}. We find that the algorithm is able to use $\zeta$ as a metric to zero out the inserted frequency drift (Figure~\ref{fig:onebias}d). The added offset bias to the purple qubit follows the inserted bias, while the compensation for qubits without an inserted bias stays near zero or a constant value. 

Importantly, we see that the $\zeta$ are stabilized in Figure~\ref{fig:onebias}e; the purple qubit stays at an average $\zeta$ of 0.12, and stays well below the randomization limit of 0.5. Thus, we show that ADEPT can be used to keep parameters for a single qubit near their ideal values. We note that the average $\zeta$ is slightly higher than the initial $\zeta$ of 0.11. This highlights a system tradeoff: in order to track the optimal value of a parameter we must sample away from the optimum of that parameter. We argue that paying this small price in error given that a future fault-tolerant computer should operate safely below threshold, and will provide a benefit in stability for long computations.

\begin{figure}[!]
    \centering
    \includegraphics[width=0.48\textwidth]{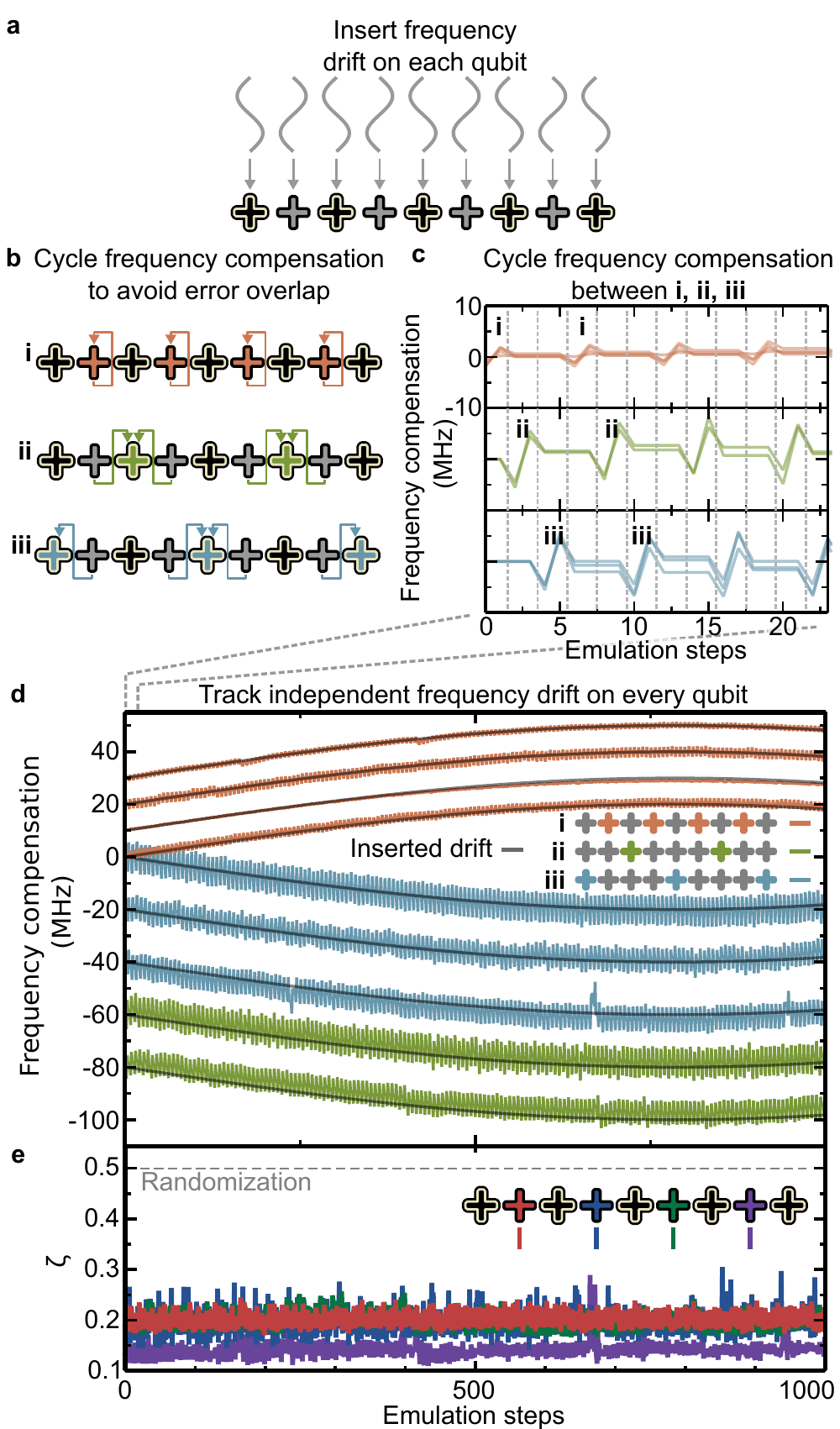}
    \caption{(Color) \textbf{Tracking and negating independent frequency drift on all qubits.} (a) A different amplitude frequency drift is inserted onto each of nine qubits in the device. (b) ADEPT is used in conjunction with a frequency following algorithm on all qubits by interleaving the active hardware pattern. In \textbf{i}, $\zeta$ is used to inform the frequency following algorithm for the measurement qubits. In \textbf{ii} and \textbf{iii} $\zeta$ is used to inform the frequency following algorithm of neighboring data qubits. (c) Compensation algorithms associated with each hardware pattern \textbf{i}-\textbf{iii} are active one pattern a time in a sequential repeating fashion. (d) Independent frequency drift on each qubit is tracked independently. The traces have been spaced along the y-axis for viewing clarity. Each data point is 6000 instances of eight rounds of detection. (e) $\zeta$ is stabilized throughout the course of the experiment, indicated that all qubits have their frequency drift compensated for.}
    \label{fig:allbias}
\end{figure}

% Julian/Qubit/9Xmon/150823/ [365, 363]
% 3000 stats, 8 cycles, 2 averages

In Figure~\ref{fig:allbias}, we show that ADEPT can be used to compensate for individual parameter drift on every qubit simultaneously. To emulate what may happen in a real system, we insert frequency drifts for each qubit of $\pm$10~MHz over 48~million emulated rounds of detection (equivalent to 42~seconds of non-emulated operation). We emphasize that this is a very large and fast drift compared to what is typically seen in hardware~\cite{bialczak20071}, making it an excellent stress test of our method. To compensate for independent drifts on each qubit, we cycle through the three different patterns, where in each pattern we optimize a subset of the qubits (Figure~\ref{fig:allbias}b). In pattern \textbf{i}, we use $\zeta$ of a measurement qubit as a metric its own frequency. In pattern \textbf{ii} and \textbf{iii} we optimize data qubits by using the average $\zeta$ of neighboring measurement qubits as a metric. By only adjusting the parameters relevant for each pattern at a time (Fig.~\ref{fig:allbias}c), we can compensate for an independent bias for each qubit (Fig.~\ref{fig:allbias}d). Using this strategy, we stabilize $\zeta$ of all measurement qubits -- indicating that \textit{all} qubits are adequately compensated for drift -- and keep them well below the randomization limit (Fig.~\ref{fig:allbias}e). This demonstrates how ADEPT can be used to keep parameters near their optimum while running long algorithms. 

We have introduced ADEPT in a one-dimensional chain of qubits running the repetition code, but this technique is generalizable to most error correction schemes. Any scheme that detects errors using groups of qubits of fixed maximum size, and the number of groups that any qubit belongs to does not scale with system size can use ADEPT. For example, this is compatible with all topological codes \cite{Brav98,Denn02,Raus07,Raus07d,Fowl12f,Bomb06,Ohze09b,Katz10,Bomb15} including subsystem codes \cite{Andr12}, and all concatenated codes \cite{Shor95,Cald95,Stea96,Knil04c,Baco06} by focusing on the lowest level of concatenation. This includes surface \cite{Fowl12f} and color codes \cite{Bomb06}, and the Steane \cite{Stea96} and Shor \cite{Shor95} codes. ADEPT may not be compatible with finite rate block codes \cite{Gott13} if one wants to preserve $O(1)$ scaling with system size. 

We have demonstrated the ADEPT control technique, which uses the error detection outcomes of operator measurements for system optimization. We have shown that this can be used to optimize gate parameters in a way that is $O(1)$ scalable to arbitrary numbers of qubits. Additionally, we have demonstrated that ADEPT can be used to compensate for system drift. By interleaving multiple hardware patterns, we can track parameter drift on every qubit, and potentially every parameter of every gate on every qubit, all without interruption of the error detection. With ADPET, we are optimistic about controlling the many physical qubits that constitute a future fault-tolerant quantum computer.

\smallskip\noindent

\smallskip\noindent
\textbf{Acknowledgements}
Devices were made at the UC Santa Barbara Nanofabrication Facility, a part of the US
NSF-funded National Nanotechnology Infrastructure Network, and at the
NanoStructures Cleanroom Facility.

\smallskip\noindent
\textbf{Author Information}
Correspondence and requests for materials should be addressed to J.K. (juliankelly@google.com)
or J.M.M. (jmartinis@google.com).

\addcontentsline{toc}{chapter}{Bibliography}
\bibliographystyle{apsrev}
\bibliography{bibliography}

\begin{thebibliography}{48}
\expandafter\ifx\csname natexlab\endcsname\relax\def\natexlab#1{#1}\fi
\expandafter\ifx\csname bibnamefont\endcsname\relax
  \def\bibnamefont#1{#1}\fi
\expandafter\ifx\csname bibfnamefont\endcsname\relax
  \def\bibfnamefont#1{#1}\fi
\expandafter\ifx\csname citenamefont\endcsname\relax
  \def\citenamefont#1{#1}\fi
\expandafter\ifx\csname url\endcsname\relax
  \def\url#1{\texttt{#1}}\fi
\expandafter\ifx\csname urlprefix\endcsname\relax\def\urlprefix{URL }\fi
\providecommand{\bibinfo}[2]{#2}
\providecommand{\eprint}[2][]{\url{#2}}

\bibitem[{\citenamefont{Fowler et~al.}(2012{\natexlab{a}})\citenamefont{Fowler,
  Mariantoni, Martinis, and Cleland}}]{fowler2012surface}
\bibinfo{author}{\bibfnamefont{A.~G.} \bibnamefont{Fowler}},
  \bibinfo{author}{\bibfnamefont{M.}~\bibnamefont{Mariantoni}},
  \bibinfo{author}{\bibfnamefont{J.~M.} \bibnamefont{Martinis}},
  \bibnamefont{and} \bibinfo{author}{\bibfnamefont{A.~N.}
  \bibnamefont{Cleland}}, \bibinfo{journal}{Phys. Rev. A}
  \textbf{\bibinfo{volume}{86}}, \bibinfo{pages}{032324}
  (\bibinfo{year}{2012}{\natexlab{a}}).

\bibitem[{\citenamefont{Chow et~al.}(2012)\citenamefont{Chow, Gambetta,
  C{\'o}rcoles, Merkel, Smolin, Rigetti, Poletto, Keefe, Rothwell, Rozen
  et~al.}}]{chow2012universal}
\bibinfo{author}{\bibfnamefont{J.~M.} \bibnamefont{Chow}},
  \bibinfo{author}{\bibfnamefont{J.~M.} \bibnamefont{Gambetta}},
  \bibinfo{author}{\bibfnamefont{A.}~\bibnamefont{C{\'o}rcoles}},
  \bibinfo{author}{\bibfnamefont{S.~T.} \bibnamefont{Merkel}},
  \bibinfo{author}{\bibfnamefont{J.~A.} \bibnamefont{Smolin}},
  \bibinfo{author}{\bibfnamefont{C.}~\bibnamefont{Rigetti}},
  \bibinfo{author}{\bibfnamefont{S.}~\bibnamefont{Poletto}},
  \bibinfo{author}{\bibfnamefont{G.~A.} \bibnamefont{Keefe}},
  \bibinfo{author}{\bibfnamefont{M.~B.} \bibnamefont{Rothwell}},
  \bibinfo{author}{\bibfnamefont{J.}~\bibnamefont{Rozen}},
  \bibnamefont{et~al.}, \bibinfo{journal}{Physical review letters}
  \textbf{\bibinfo{volume}{109}}, \bibinfo{pages}{060501}
  (\bibinfo{year}{2012}).

\bibitem[{\citenamefont{Harty et~al.}(2014)\citenamefont{Harty, Allcock,
  Ballance, Guidoni, Janacek, Linke, Stacey, and Lucas}}]{harty2014high}
\bibinfo{author}{\bibfnamefont{T.}~\bibnamefont{Harty}},
  \bibinfo{author}{\bibfnamefont{D.}~\bibnamefont{Allcock}},
  \bibinfo{author}{\bibfnamefont{C.}~\bibnamefont{Ballance}},
  \bibinfo{author}{\bibfnamefont{L.}~\bibnamefont{Guidoni}},
  \bibinfo{author}{\bibfnamefont{H.}~\bibnamefont{Janacek}},
  \bibinfo{author}{\bibfnamefont{N.}~\bibnamefont{Linke}},
  \bibinfo{author}{\bibfnamefont{D.}~\bibnamefont{Stacey}}, \bibnamefont{and}
  \bibinfo{author}{\bibfnamefont{D.}~\bibnamefont{Lucas}},
  \bibinfo{journal}{Physical review letters} \textbf{\bibinfo{volume}{113}},
  \bibinfo{pages}{220501} (\bibinfo{year}{2014}).

\bibitem[{\citenamefont{Barends et~al.}(2014)\citenamefont{Barends, Kelly,
  Megrant, Veitia, Sank, Jeffrey, White, Mutus, Fowler, Campbell
  et~al.}}]{barends2014superconducting}
\bibinfo{author}{\bibfnamefont{R.}~\bibnamefont{Barends}},
  \bibinfo{author}{\bibfnamefont{J.}~\bibnamefont{Kelly}},
  \bibinfo{author}{\bibfnamefont{A.}~\bibnamefont{Megrant}},
  \bibinfo{author}{\bibfnamefont{A.}~\bibnamefont{Veitia}},
  \bibinfo{author}{\bibfnamefont{D.}~\bibnamefont{Sank}},
  \bibinfo{author}{\bibfnamefont{E.}~\bibnamefont{Jeffrey}},
  \bibinfo{author}{\bibfnamefont{T.}~\bibnamefont{White}},
  \bibinfo{author}{\bibfnamefont{J.}~\bibnamefont{Mutus}},
  \bibinfo{author}{\bibfnamefont{A.}~\bibnamefont{Fowler}},
  \bibinfo{author}{\bibfnamefont{B.}~\bibnamefont{Campbell}},
  \bibnamefont{et~al.}, \bibinfo{journal}{Nature}
  \textbf{\bibinfo{volume}{508}}, \bibinfo{pages}{500} (\bibinfo{year}{2014}).

\bibitem[{\citenamefont{Ballance et~al.}(2015)\citenamefont{Ballance, Harty,
  Linke, Sepiol, and Lucas}}]{ballance2015laser}
\bibinfo{author}{\bibfnamefont{C.}~\bibnamefont{Ballance}},
  \bibinfo{author}{\bibfnamefont{T.}~\bibnamefont{Harty}},
  \bibinfo{author}{\bibfnamefont{N.}~\bibnamefont{Linke}},
  \bibinfo{author}{\bibfnamefont{M.}~\bibnamefont{Sepiol}}, \bibnamefont{and}
  \bibinfo{author}{\bibfnamefont{D.}~\bibnamefont{Lucas}},
  \bibinfo{journal}{arXiv preprint arXiv:1512.04600}  (\bibinfo{year}{2015}).

\bibitem[{\citenamefont{Kelly et~al.}(2015)\citenamefont{Kelly, Barends,
  Fowler, Megrant, Jeffrey, White, Sank, Mutus, Campbell, Chen
  et~al.}}]{kelly2015state}
\bibinfo{author}{\bibfnamefont{J.}~\bibnamefont{Kelly}},
  \bibinfo{author}{\bibfnamefont{R.}~\bibnamefont{Barends}},
  \bibinfo{author}{\bibfnamefont{A.}~\bibnamefont{Fowler}},
  \bibinfo{author}{\bibfnamefont{A.}~\bibnamefont{Megrant}},
  \bibinfo{author}{\bibfnamefont{E.}~\bibnamefont{Jeffrey}},
  \bibinfo{author}{\bibfnamefont{T.}~\bibnamefont{White}},
  \bibinfo{author}{\bibfnamefont{D.}~\bibnamefont{Sank}},
  \bibinfo{author}{\bibfnamefont{J.}~\bibnamefont{Mutus}},
  \bibinfo{author}{\bibfnamefont{B.}~\bibnamefont{Campbell}},
  \bibinfo{author}{\bibfnamefont{Y.}~\bibnamefont{Chen}}, \bibnamefont{et~al.},
  \bibinfo{journal}{Nature} \textbf{\bibinfo{volume}{519}}, \bibinfo{pages}{66}
  (\bibinfo{year}{2015}).

\bibitem[{\citenamefont{C{\'o}rcoles et~al.}(2015)\citenamefont{C{\'o}rcoles,
  Magesan, Srinivasan, Cross, Steffen, Gambetta, and
  Chow}}]{corcoles2015demonstration}
\bibinfo{author}{\bibfnamefont{A.}~\bibnamefont{C{\'o}rcoles}},
  \bibinfo{author}{\bibfnamefont{E.}~\bibnamefont{Magesan}},
  \bibinfo{author}{\bibfnamefont{S.~J.} \bibnamefont{Srinivasan}},
  \bibinfo{author}{\bibfnamefont{A.~W.} \bibnamefont{Cross}},
  \bibinfo{author}{\bibfnamefont{M.}~\bibnamefont{Steffen}},
  \bibinfo{author}{\bibfnamefont{J.~M.} \bibnamefont{Gambetta}},
  \bibnamefont{and} \bibinfo{author}{\bibfnamefont{J.~M.} \bibnamefont{Chow}},
  \bibinfo{journal}{Nature communications} \textbf{\bibinfo{volume}{6}}
  (\bibinfo{year}{2015}).

\bibitem[{\citenamefont{Chow et~al.}(2014)\citenamefont{Chow, Gambetta,
  Magesan, Abraham, Cross, Johnson, Masluk, Ryan, Smolin, Srinivasan
  et~al.}}]{chow2014implementing}
\bibinfo{author}{\bibfnamefont{J.~M.} \bibnamefont{Chow}},
  \bibinfo{author}{\bibfnamefont{J.~M.} \bibnamefont{Gambetta}},
  \bibinfo{author}{\bibfnamefont{E.}~\bibnamefont{Magesan}},
  \bibinfo{author}{\bibfnamefont{D.~W.} \bibnamefont{Abraham}},
  \bibinfo{author}{\bibfnamefont{A.~W.} \bibnamefont{Cross}},
  \bibinfo{author}{\bibfnamefont{B.}~\bibnamefont{Johnson}},
  \bibinfo{author}{\bibfnamefont{N.~A.} \bibnamefont{Masluk}},
  \bibinfo{author}{\bibfnamefont{C.~A.} \bibnamefont{Ryan}},
  \bibinfo{author}{\bibfnamefont{J.~A.} \bibnamefont{Smolin}},
  \bibinfo{author}{\bibfnamefont{S.~J.} \bibnamefont{Srinivasan}},
  \bibnamefont{et~al.}, \bibinfo{journal}{Nature communications}
  \textbf{\bibinfo{volume}{5}} (\bibinfo{year}{2014}).

\bibitem[{\citenamefont{Gambetta et~al.}(2015)\citenamefont{Gambetta, Chow, and
  Steffen}}]{gambetta2015building}
\bibinfo{author}{\bibfnamefont{J.~M.} \bibnamefont{Gambetta}},
  \bibinfo{author}{\bibfnamefont{J.~M.} \bibnamefont{Chow}}, \bibnamefont{and}
  \bibinfo{author}{\bibfnamefont{M.}~\bibnamefont{Steffen}},
  \bibinfo{journal}{arXiv preprint arXiv:1510.04375}  (\bibinfo{year}{2015}).

\bibitem[{\citenamefont{Bialczak et~al.}(2007)\citenamefont{Bialczak,
  McDermott, Ansmann, Hofheinz, Katz, Lucero, Neeley, O'Connell, Wang, Cleland
  et~al.}}]{bialczak20071}
\bibinfo{author}{\bibfnamefont{R.~C.} \bibnamefont{Bialczak}},
  \bibinfo{author}{\bibfnamefont{R.}~\bibnamefont{McDermott}},
  \bibinfo{author}{\bibfnamefont{M.}~\bibnamefont{Ansmann}},
  \bibinfo{author}{\bibfnamefont{M.}~\bibnamefont{Hofheinz}},
  \bibinfo{author}{\bibfnamefont{N.}~\bibnamefont{Katz}},
  \bibinfo{author}{\bibfnamefont{E.}~\bibnamefont{Lucero}},
  \bibinfo{author}{\bibfnamefont{M.}~\bibnamefont{Neeley}},
  \bibinfo{author}{\bibfnamefont{A.}~\bibnamefont{O'Connell}},
  \bibinfo{author}{\bibfnamefont{H.}~\bibnamefont{Wang}},
  \bibinfo{author}{\bibfnamefont{A.}~\bibnamefont{Cleland}},
  \bibnamefont{et~al.}, \bibinfo{journal}{Physical review letters}
  \textbf{\bibinfo{volume}{99}}, \bibinfo{pages}{187006}
  (\bibinfo{year}{2007}).

\bibitem[{\citenamefont{Miquel et~al.}(1997)\citenamefont{Miquel, Paz, and
  Zurek}}]{miquel1997quantum}
\bibinfo{author}{\bibfnamefont{C.}~\bibnamefont{Miquel}},
  \bibinfo{author}{\bibfnamefont{J.~P.} \bibnamefont{Paz}}, \bibnamefont{and}
  \bibinfo{author}{\bibfnamefont{W.~H.} \bibnamefont{Zurek}},
  \bibinfo{journal}{Physical review letters} \textbf{\bibinfo{volume}{78}},
  \bibinfo{pages}{3971} (\bibinfo{year}{1997}).

\bibitem[{\citenamefont{Shapiro and Brumer}(2003)}]{shapiro2003principles}
\bibinfo{author}{\bibfnamefont{M.}~\bibnamefont{Shapiro}} \bibnamefont{and}
  \bibinfo{author}{\bibfnamefont{P.}~\bibnamefont{Brumer}},
  \bibinfo{journal}{Principles of the Quantum Control of Molecular Processes,
  by Moshe Shapiro, Paul Brumer, pp. 250. ISBN 0-471-24184-9. Wiley-VCH,
  February 2003.} \textbf{\bibinfo{volume}{1}} (\bibinfo{year}{2003}).

\bibitem[{\citenamefont{Khaneja et~al.}(2005)\citenamefont{Khaneja, Reiss,
  Kehlet, Schulte-Herbr{\"u}ggen, and Glaser}}]{khaneja2005optimal}
\bibinfo{author}{\bibfnamefont{N.}~\bibnamefont{Khaneja}},
  \bibinfo{author}{\bibfnamefont{T.}~\bibnamefont{Reiss}},
  \bibinfo{author}{\bibfnamefont{C.}~\bibnamefont{Kehlet}},
  \bibinfo{author}{\bibfnamefont{T.}~\bibnamefont{Schulte-Herbr{\"u}ggen}},
  \bibnamefont{and} \bibinfo{author}{\bibfnamefont{S.~J.}
  \bibnamefont{Glaser}}, \bibinfo{journal}{Journal of Magnetic Resonance}
  \textbf{\bibinfo{volume}{172}}, \bibinfo{pages}{296} (\bibinfo{year}{2005}).

\bibitem[{\citenamefont{Bardeen et~al.}(1997)\citenamefont{Bardeen, Yakovlev,
  Wilson, Carpenter, Weber, and Warren}}]{bardeen1997feedback}
\bibinfo{author}{\bibfnamefont{C.~J.} \bibnamefont{Bardeen}},
  \bibinfo{author}{\bibfnamefont{V.~V.} \bibnamefont{Yakovlev}},
  \bibinfo{author}{\bibfnamefont{K.~R.} \bibnamefont{Wilson}},
  \bibinfo{author}{\bibfnamefont{S.~D.} \bibnamefont{Carpenter}},
  \bibinfo{author}{\bibfnamefont{P.~M.} \bibnamefont{Weber}}, \bibnamefont{and}
  \bibinfo{author}{\bibfnamefont{W.~S.} \bibnamefont{Warren}},
  \bibinfo{journal}{Chemical Physics Letters} \textbf{\bibinfo{volume}{280}},
  \bibinfo{pages}{151} (\bibinfo{year}{1997}).

\bibitem[{\citenamefont{Koch et~al.}(2004)\citenamefont{Koch, Palao, Kosloff,
  and Masnou-Seeuws}}]{koch2004stabilization}
\bibinfo{author}{\bibfnamefont{C.~P.} \bibnamefont{Koch}},
  \bibinfo{author}{\bibfnamefont{J.~P.} \bibnamefont{Palao}},
  \bibinfo{author}{\bibfnamefont{R.}~\bibnamefont{Kosloff}}, \bibnamefont{and}
  \bibinfo{author}{\bibfnamefont{F.}~\bibnamefont{Masnou-Seeuws}},
  \bibinfo{journal}{Physical Review A} \textbf{\bibinfo{volume}{70}},
  \bibinfo{pages}{013402} (\bibinfo{year}{2004}).

\bibitem[{\citenamefont{Hohenester et~al.}(2007)\citenamefont{Hohenester,
  Rekdal, Borz{\`\i}, and Schmiedmayer}}]{hohenester2007optimal}
\bibinfo{author}{\bibfnamefont{U.}~\bibnamefont{Hohenester}},
  \bibinfo{author}{\bibfnamefont{P.~K.} \bibnamefont{Rekdal}},
  \bibinfo{author}{\bibfnamefont{A.}~\bibnamefont{Borz{\`\i}}},
  \bibnamefont{and}
  \bibinfo{author}{\bibfnamefont{J.}~\bibnamefont{Schmiedmayer}},
  \bibinfo{journal}{Physical Review A} \textbf{\bibinfo{volume}{75}},
  \bibinfo{pages}{023602} (\bibinfo{year}{2007}).

\bibitem[{\citenamefont{Sp{\"o}rl et~al.}(2007)\citenamefont{Sp{\"o}rl,
  Schulte-Herbr{\"u}ggen, Glaser, Bergholm, Storcz, Ferber, and
  Wilhelm}}]{sporl2007optimal}
\bibinfo{author}{\bibfnamefont{A.}~\bibnamefont{Sp{\"o}rl}},
  \bibinfo{author}{\bibfnamefont{T.}~\bibnamefont{Schulte-Herbr{\"u}ggen}},
  \bibinfo{author}{\bibfnamefont{S.}~\bibnamefont{Glaser}},
  \bibinfo{author}{\bibfnamefont{V.}~\bibnamefont{Bergholm}},
  \bibinfo{author}{\bibfnamefont{M.}~\bibnamefont{Storcz}},
  \bibinfo{author}{\bibfnamefont{J.}~\bibnamefont{Ferber}}, \bibnamefont{and}
  \bibinfo{author}{\bibfnamefont{F.}~\bibnamefont{Wilhelm}},
  \bibinfo{journal}{Physical Review A} \textbf{\bibinfo{volume}{75}},
  \bibinfo{pages}{012302} (\bibinfo{year}{2007}).

\bibitem[{\citenamefont{Kelly et~al.}(2014)\citenamefont{Kelly, Barends,
  Campbell, Chen, Chen, Chiaro, Dunsworth, Fowler, Hoi, Jeffrey
  et~al.}}]{kelly2014optimal}
\bibinfo{author}{\bibfnamefont{J.}~\bibnamefont{Kelly}},
  \bibinfo{author}{\bibfnamefont{R.}~\bibnamefont{Barends}},
  \bibinfo{author}{\bibfnamefont{B.}~\bibnamefont{Campbell}},
  \bibinfo{author}{\bibfnamefont{Y.}~\bibnamefont{Chen}},
  \bibinfo{author}{\bibfnamefont{Z.}~\bibnamefont{Chen}},
  \bibinfo{author}{\bibfnamefont{B.}~\bibnamefont{Chiaro}},
  \bibinfo{author}{\bibfnamefont{A.}~\bibnamefont{Dunsworth}},
  \bibinfo{author}{\bibfnamefont{A.~G.} \bibnamefont{Fowler}},
  \bibinfo{author}{\bibfnamefont{I.-C.} \bibnamefont{Hoi}},
  \bibinfo{author}{\bibfnamefont{E.}~\bibnamefont{Jeffrey}},
  \bibnamefont{et~al.}, \bibinfo{journal}{Phys. Rev. Lett.}
  \textbf{\bibinfo{volume}{112}}, \bibinfo{pages}{240504}
  (\bibinfo{year}{2014}).

\bibitem[{\citenamefont{Egger and
  Wilhelm}(2014{\natexlab{a}})}]{egger2014adaptive}
\bibinfo{author}{\bibfnamefont{D.}~\bibnamefont{Egger}} \bibnamefont{and}
  \bibinfo{author}{\bibfnamefont{F.}~\bibnamefont{Wilhelm}},
  \bibinfo{journal}{Physical review letters} \textbf{\bibinfo{volume}{112}},
  \bibinfo{pages}{240503} (\bibinfo{year}{2014}{\natexlab{a}}).

\bibitem[{\citenamefont{Egger and
  Wilhelm}(2014{\natexlab{b}})}]{egger2014optimized}
\bibinfo{author}{\bibfnamefont{D.~J.} \bibnamefont{Egger}} \bibnamefont{and}
  \bibinfo{author}{\bibfnamefont{F.~K.} \bibnamefont{Wilhelm}},
  \bibinfo{journal}{Superconductor Science and Technology}
  \textbf{\bibinfo{volume}{27}}, \bibinfo{pages}{014001}
  (\bibinfo{year}{2014}{\natexlab{b}}).

\bibitem[{\citenamefont{Zhang et~al.}(2011{\natexlab{a}})\citenamefont{Zhang,
  Ryan, Laflamme, and Baugh}}]{zhang2011coherent}
\bibinfo{author}{\bibfnamefont{Y.}~\bibnamefont{Zhang}},
  \bibinfo{author}{\bibfnamefont{C.~A.} \bibnamefont{Ryan}},
  \bibinfo{author}{\bibfnamefont{R.}~\bibnamefont{Laflamme}}, \bibnamefont{and}
  \bibinfo{author}{\bibfnamefont{J.}~\bibnamefont{Baugh}},
  \bibinfo{journal}{Physical review letters} \textbf{\bibinfo{volume}{107}},
  \bibinfo{pages}{170503} (\bibinfo{year}{2011}{\natexlab{a}}).

\bibitem[{\citenamefont{Zhang et~al.}(2011{\natexlab{b}})\citenamefont{Zhang,
  Gangloff, Moussa, and Laflamme}}]{zhang2011experimental}
\bibinfo{author}{\bibfnamefont{J.}~\bibnamefont{Zhang}},
  \bibinfo{author}{\bibfnamefont{D.}~\bibnamefont{Gangloff}},
  \bibinfo{author}{\bibfnamefont{O.}~\bibnamefont{Moussa}}, \bibnamefont{and}
  \bibinfo{author}{\bibfnamefont{R.}~\bibnamefont{Laflamme}},
  \bibinfo{journal}{Physical Review A} \textbf{\bibinfo{volume}{84}},
  \bibinfo{pages}{034303} (\bibinfo{year}{2011}{\natexlab{b}}).

\bibitem[{\citenamefont{O'Brien et~al.}(2004)\citenamefont{O'Brien, Pryde,
  Gilchrist, James, Langford, Ralph, and White}}]{o2004quantum}
\bibinfo{author}{\bibfnamefont{J.~L.} \bibnamefont{O'Brien}},
  \bibinfo{author}{\bibfnamefont{G.}~\bibnamefont{Pryde}},
  \bibinfo{author}{\bibfnamefont{A.}~\bibnamefont{Gilchrist}},
  \bibinfo{author}{\bibfnamefont{D.}~\bibnamefont{James}},
  \bibinfo{author}{\bibfnamefont{N.}~\bibnamefont{Langford}},
  \bibinfo{author}{\bibfnamefont{T.}~\bibnamefont{Ralph}}, \bibnamefont{and}
  \bibinfo{author}{\bibfnamefont{A.}~\bibnamefont{White}},
  \bibinfo{journal}{Physical review letters} \textbf{\bibinfo{volume}{93}},
  \bibinfo{pages}{080502} (\bibinfo{year}{2004}).

\bibitem[{\citenamefont{Magesan et~al.}(2011)\citenamefont{Magesan, Gambetta,
  and Emerson}}]{magesan2011scalable}
\bibinfo{author}{\bibfnamefont{E.}~\bibnamefont{Magesan}},
  \bibinfo{author}{\bibfnamefont{J.~M.} \bibnamefont{Gambetta}},
  \bibnamefont{and} \bibinfo{author}{\bibfnamefont{J.}~\bibnamefont{Emerson}},
  \bibinfo{journal}{Physical review letters} \textbf{\bibinfo{volume}{106}},
  \bibinfo{pages}{180504} (\bibinfo{year}{2011}).

\bibitem[{\citenamefont{Calarco et~al.}(2004)\citenamefont{Calarco, Dorner,
  Julienne, Williams, and Zoller}}]{calarco2004quantum}
\bibinfo{author}{\bibfnamefont{T.}~\bibnamefont{Calarco}},
  \bibinfo{author}{\bibfnamefont{U.}~\bibnamefont{Dorner}},
  \bibinfo{author}{\bibfnamefont{P.}~\bibnamefont{Julienne}},
  \bibinfo{author}{\bibfnamefont{C.}~\bibnamefont{Williams}}, \bibnamefont{and}
  \bibinfo{author}{\bibfnamefont{P.}~\bibnamefont{Zoller}},
  \bibinfo{journal}{Physical Review A} \textbf{\bibinfo{volume}{70}},
  \bibinfo{pages}{012306} (\bibinfo{year}{2004}).

\bibitem[{\citenamefont{Koch et~al.}(2007)\citenamefont{Koch, Terri, Gambetta,
  Houck, Schuster, Majer, Blais, Devoret, Girvin, and
  Schoelkopf}}]{koch2007charge}
\bibinfo{author}{\bibfnamefont{J.}~\bibnamefont{Koch}},
  \bibinfo{author}{\bibfnamefont{M.~Y.} \bibnamefont{Terri}},
  \bibinfo{author}{\bibfnamefont{J.}~\bibnamefont{Gambetta}},
  \bibinfo{author}{\bibfnamefont{A.~A.} \bibnamefont{Houck}},
  \bibinfo{author}{\bibfnamefont{D.}~\bibnamefont{Schuster}},
  \bibinfo{author}{\bibfnamefont{J.}~\bibnamefont{Majer}},
  \bibinfo{author}{\bibfnamefont{A.}~\bibnamefont{Blais}},
  \bibinfo{author}{\bibfnamefont{M.~H.} \bibnamefont{Devoret}},
  \bibinfo{author}{\bibfnamefont{S.~M.} \bibnamefont{Girvin}},
  \bibnamefont{and} \bibinfo{author}{\bibfnamefont{R.~J.}
  \bibnamefont{Schoelkopf}}, \bibinfo{journal}{Physical Review A}
  \textbf{\bibinfo{volume}{76}}, \bibinfo{pages}{042319}
  (\bibinfo{year}{2007}).

\bibitem[{\citenamefont{Barends et~al.}(2013)\citenamefont{Barends, Kelly,
  Megrant, Sank, Jeffrey, Chen, Yin, Chiaro, Mutus, Neill
  et~al.}}]{barends2013coherent}
\bibinfo{author}{\bibfnamefont{R.}~\bibnamefont{Barends}},
  \bibinfo{author}{\bibfnamefont{J.}~\bibnamefont{Kelly}},
  \bibinfo{author}{\bibfnamefont{A.}~\bibnamefont{Megrant}},
  \bibinfo{author}{\bibfnamefont{D.}~\bibnamefont{Sank}},
  \bibinfo{author}{\bibfnamefont{E.}~\bibnamefont{Jeffrey}},
  \bibinfo{author}{\bibfnamefont{Y.}~\bibnamefont{Chen}},
  \bibinfo{author}{\bibfnamefont{Y.}~\bibnamefont{Yin}},
  \bibinfo{author}{\bibfnamefont{B.}~\bibnamefont{Chiaro}},
  \bibinfo{author}{\bibfnamefont{J.}~\bibnamefont{Mutus}},
  \bibinfo{author}{\bibfnamefont{C.}~\bibnamefont{Neill}},
  \bibnamefont{et~al.}, \bibinfo{journal}{Phys. Rev. Lett.}
  \textbf{\bibinfo{volume}{111}}, \bibinfo{pages}{080502}
  (\bibinfo{year}{2013}).

\bibitem[{\citenamefont{Schuster et~al.}(2005)\citenamefont{Schuster, Wallraff,
  Blais, Frunzio, Huang, Majer, Girvin, and Schoelkopf}}]{schuster2005ac}
\bibinfo{author}{\bibfnamefont{D.}~\bibnamefont{Schuster}},
  \bibinfo{author}{\bibfnamefont{A.}~\bibnamefont{Wallraff}},
  \bibinfo{author}{\bibfnamefont{A.}~\bibnamefont{Blais}},
  \bibinfo{author}{\bibfnamefont{L.}~\bibnamefont{Frunzio}},
  \bibinfo{author}{\bibfnamefont{R.-S.} \bibnamefont{Huang}},
  \bibinfo{author}{\bibfnamefont{J.}~\bibnamefont{Majer}},
  \bibinfo{author}{\bibfnamefont{S.}~\bibnamefont{Girvin}}, \bibnamefont{and}
  \bibinfo{author}{\bibfnamefont{R.}~\bibnamefont{Schoelkopf}},
  \bibinfo{journal}{Physical review letters} \textbf{\bibinfo{volume}{94}},
  \bibinfo{pages}{123602} (\bibinfo{year}{2005}).

\bibitem[{\citenamefont{Wallraff et~al.}(2005)\citenamefont{Wallraff, Schuster,
  Blais, Frunzio, Majer, Devoret, Girvin, and
  Schoelkopf}}]{wallraff2005approaching}
\bibinfo{author}{\bibfnamefont{A.}~\bibnamefont{Wallraff}},
  \bibinfo{author}{\bibfnamefont{D.}~\bibnamefont{Schuster}},
  \bibinfo{author}{\bibfnamefont{A.}~\bibnamefont{Blais}},
  \bibinfo{author}{\bibfnamefont{L.}~\bibnamefont{Frunzio}},
  \bibinfo{author}{\bibfnamefont{J.}~\bibnamefont{Majer}},
  \bibinfo{author}{\bibfnamefont{M.}~\bibnamefont{Devoret}},
  \bibinfo{author}{\bibfnamefont{S.}~\bibnamefont{Girvin}}, \bibnamefont{and}
  \bibinfo{author}{\bibfnamefont{R.}~\bibnamefont{Schoelkopf}},
  \bibinfo{journal}{Physical Review Letters} \textbf{\bibinfo{volume}{95}},
  \bibinfo{pages}{060501} (\bibinfo{year}{2005}).

\bibitem[{\citenamefont{Jeffrey et~al.}(2014)\citenamefont{Jeffrey, Sank,
  Mutus, White, Kelly, Barends, Chen, Chen, Chiaro, Dunsworth
  et~al.}}]{jeffrey2014fast}
\bibinfo{author}{\bibfnamefont{E.}~\bibnamefont{Jeffrey}},
  \bibinfo{author}{\bibfnamefont{D.}~\bibnamefont{Sank}},
  \bibinfo{author}{\bibfnamefont{J.~Y.} \bibnamefont{Mutus}},
  \bibinfo{author}{\bibfnamefont{T.~C.} \bibnamefont{White}},
  \bibinfo{author}{\bibfnamefont{J.}~\bibnamefont{Kelly}},
  \bibinfo{author}{\bibfnamefont{R.}~\bibnamefont{Barends}},
  \bibinfo{author}{\bibfnamefont{Y.}~\bibnamefont{Chen}},
  \bibinfo{author}{\bibfnamefont{Z.}~\bibnamefont{Chen}},
  \bibinfo{author}{\bibfnamefont{B.}~\bibnamefont{Chiaro}},
  \bibinfo{author}{\bibfnamefont{A.}~\bibnamefont{Dunsworth}},
  \bibnamefont{et~al.}, \bibinfo{journal}{Phys. Rev. Lett.}
  \textbf{\bibinfo{volume}{112}}, \bibinfo{pages}{190504}
  (\bibinfo{year}{2014}).

\bibitem[{\citenamefont{Mutus et~al.}(2014)\citenamefont{Mutus, White, Barends,
  Chen, Chen, Chiaro, Dunsworth, Jeffrey, Kelly, Megrant
  et~al.}}]{mutus2014strong}
\bibinfo{author}{\bibfnamefont{J.~Y.} \bibnamefont{Mutus}},
  \bibinfo{author}{\bibfnamefont{T.~C.} \bibnamefont{White}},
  \bibinfo{author}{\bibfnamefont{R.}~\bibnamefont{Barends}},
  \bibinfo{author}{\bibfnamefont{Y.}~\bibnamefont{Chen}},
  \bibinfo{author}{\bibfnamefont{Z.}~\bibnamefont{Chen}},
  \bibinfo{author}{\bibfnamefont{B.}~\bibnamefont{Chiaro}},
  \bibinfo{author}{\bibfnamefont{A.}~\bibnamefont{Dunsworth}},
  \bibinfo{author}{\bibfnamefont{E.}~\bibnamefont{Jeffrey}},
  \bibinfo{author}{\bibfnamefont{J.}~\bibnamefont{Kelly}},
  \bibinfo{author}{\bibfnamefont{A.}~\bibnamefont{Megrant}},
  \bibnamefont{et~al.}, \bibinfo{journal}{Applied Physics Letters}
  \textbf{\bibinfo{volume}{104}}, \bibinfo{eid}{263513} (\bibinfo{year}{2014}).

\bibitem[{sup()}]{supp}
\bibinfo{note}{See supplementary information for discussion of ADEPT hardware
  patterns, emulating continuously running error detection, and details of the
  bias tracking algorithm.}

\bibitem[{\citenamefont{Bravyi and Kitaev}(1998)}]{Brav98}
\bibinfo{author}{\bibfnamefont{S.~B.} \bibnamefont{Bravyi}} \bibnamefont{and}
  \bibinfo{author}{\bibfnamefont{A.~Y.} \bibnamefont{Kitaev}},
  \bibinfo{journal}{quant-ph/9811052}  (\bibinfo{year}{1998}).

\bibitem[{\citenamefont{Dennis et~al.}(2002)\citenamefont{Dennis, Kitaev,
  Landahl, and Preskill}}]{Denn02}
\bibinfo{author}{\bibfnamefont{E.}~\bibnamefont{Dennis}},
  \bibinfo{author}{\bibfnamefont{A.}~\bibnamefont{Kitaev}},
  \bibinfo{author}{\bibfnamefont{A.}~\bibnamefont{Landahl}}, \bibnamefont{and}
  \bibinfo{author}{\bibfnamefont{J.}~\bibnamefont{Preskill}},
  \bibinfo{journal}{J. Math. Phys.} \textbf{\bibinfo{volume}{43}},
  \bibinfo{pages}{4452} (\bibinfo{year}{2002}),
  \bibinfo{note}{quant-ph/0110143}.

\bibitem[{\citenamefont{Raussendorf and Harrington}(2007)}]{Raus07}
\bibinfo{author}{\bibfnamefont{R.}~\bibnamefont{Raussendorf}} \bibnamefont{and}
  \bibinfo{author}{\bibfnamefont{J.}~\bibnamefont{Harrington}},
  \bibinfo{journal}{Phys. Rev. Lett.} \textbf{\bibinfo{volume}{98}},
  \bibinfo{pages}{190504} (\bibinfo{year}{2007}),
  \bibinfo{note}{quant-ph/0610082}.

\bibitem[{\citenamefont{Raussendorf et~al.}(2007)\citenamefont{Raussendorf,
  Harrington, and Goyal}}]{Raus07d}
\bibinfo{author}{\bibfnamefont{R.}~\bibnamefont{Raussendorf}},
  \bibinfo{author}{\bibfnamefont{J.}~\bibnamefont{Harrington}},
  \bibnamefont{and} \bibinfo{author}{\bibfnamefont{K.}~\bibnamefont{Goyal}},
  \bibinfo{journal}{New J. Phys.} \textbf{\bibinfo{volume}{9}},
  \bibinfo{pages}{199} (\bibinfo{year}{2007}),
  \bibinfo{note}{quant-ph/0703143}.

\bibitem[{\citenamefont{Fowler et~al.}(2012{\natexlab{b}})\citenamefont{Fowler,
  Mariantoni, Martinis, and Cleland}}]{Fowl12f}
\bibinfo{author}{\bibfnamefont{A.~G.} \bibnamefont{Fowler}},
  \bibinfo{author}{\bibfnamefont{M.}~\bibnamefont{Mariantoni}},
  \bibinfo{author}{\bibfnamefont{J.~M.} \bibnamefont{Martinis}},
  \bibnamefont{and} \bibinfo{author}{\bibfnamefont{A.~N.}
  \bibnamefont{Cleland}}, \bibinfo{journal}{Phys. Rev. A}
  \textbf{\bibinfo{volume}{86}}, \bibinfo{pages}{032324}
  (\bibinfo{year}{2012}{\natexlab{b}}), \bibinfo{note}{arXiv:1208.0928}.

\bibitem[{\citenamefont{Bombin and Martin-Delgado}(2006)}]{Bomb06}
\bibinfo{author}{\bibfnamefont{H.}~\bibnamefont{Bombin}} \bibnamefont{and}
  \bibinfo{author}{\bibfnamefont{M.~A.} \bibnamefont{Martin-Delgado}},
  \bibinfo{journal}{Phys. Rev. Lett.} \textbf{\bibinfo{volume}{97}},
  \bibinfo{pages}{180501} (\bibinfo{year}{2006}),
  \bibinfo{note}{quant-ph/0605138}.

\bibitem[{\citenamefont{Ohzeki}(2009)}]{Ohze09b}
\bibinfo{author}{\bibfnamefont{M.}~\bibnamefont{Ohzeki}},
  \bibinfo{journal}{Phys. Rev. E} \textbf{\bibinfo{volume}{80}},
  \bibinfo{pages}{011141} (\bibinfo{year}{2009}),
  \bibinfo{note}{arXiv:0903.2102}.

\bibitem[{\citenamefont{Katzgraber et~al.}(2010)\citenamefont{Katzgraber,
  Bombin, Andrist, and Martin-Delgado}}]{Katz10}
\bibinfo{author}{\bibfnamefont{H.~G.} \bibnamefont{Katzgraber}},
  \bibinfo{author}{\bibfnamefont{H.}~\bibnamefont{Bombin}},
  \bibinfo{author}{\bibfnamefont{R.~S.} \bibnamefont{Andrist}},
  \bibnamefont{and} \bibinfo{author}{\bibfnamefont{M.~A.}
  \bibnamefont{Martin-Delgado}}, \bibinfo{journal}{Phys. Rev. A}
  \textbf{\bibinfo{volume}{81}}, \bibinfo{pages}{012319}
  (\bibinfo{year}{2010}), \bibinfo{note}{arXiv:0910.0573}.

\bibitem[{\citenamefont{Bombin}(2015)}]{Bomb15}
\bibinfo{author}{\bibfnamefont{H.}~\bibnamefont{Bombin}},
  \bibinfo{journal}{Phys. Rev. X} \textbf{\bibinfo{volume}{5}},
  \bibinfo{pages}{031043} (\bibinfo{year}{2015}),
  \bibinfo{note}{arXiv:1404.5504}.

\bibitem[{\citenamefont{Andrist et~al.}(2012)\citenamefont{Andrist, Bombin,
  Katzgraber, and Martin-Delgado}}]{Andr12}
\bibinfo{author}{\bibfnamefont{R.~S.} \bibnamefont{Andrist}},
  \bibinfo{author}{\bibfnamefont{H.}~\bibnamefont{Bombin}},
  \bibinfo{author}{\bibfnamefont{H.~G.} \bibnamefont{Katzgraber}},
  \bibnamefont{and} \bibinfo{author}{\bibfnamefont{M.~A.}
  \bibnamefont{Martin-Delgado}}, \bibinfo{journal}{Phys. Rev. A}
  \textbf{\bibinfo{volume}{85}}, \bibinfo{pages}{050302R}
  (\bibinfo{year}{2012}), \bibinfo{note}{arXiv:1204.1838}.

\bibitem[{\citenamefont{Shor}(1995)}]{Shor95}
\bibinfo{author}{\bibfnamefont{P.~W.} \bibnamefont{Shor}},
  \bibinfo{journal}{Phys. Rev. A} \textbf{\bibinfo{volume}{52}},
  \bibinfo{pages}{R2493} (\bibinfo{year}{1995}).

\bibitem[{\citenamefont{Calderbank and Shor}(1996)}]{Cald95}
\bibinfo{author}{\bibfnamefont{A.~R.} \bibnamefont{Calderbank}}
  \bibnamefont{and} \bibinfo{author}{\bibfnamefont{P.~W.} \bibnamefont{Shor}},
  \bibinfo{journal}{Phys. Rev. A} \textbf{\bibinfo{volume}{54}},
  \bibinfo{pages}{1098} (\bibinfo{year}{1996}),
  \bibinfo{note}{quant-ph/9512032}.

\bibitem[{\citenamefont{Steane}(1996)}]{Stea96}
\bibinfo{author}{\bibfnamefont{A.~M.} \bibnamefont{Steane}},
  \bibinfo{journal}{Proc. R. Soc. Lond. A} \textbf{\bibinfo{volume}{452}},
  \bibinfo{pages}{2551} (\bibinfo{year}{1996}),
  \bibinfo{note}{quant-ph/9601029}.

\bibitem[{\citenamefont{Knill}(2005)}]{Knil04c}
\bibinfo{author}{\bibfnamefont{E.}~\bibnamefont{Knill}},
  \bibinfo{journal}{Nature} \textbf{\bibinfo{volume}{434}}, \bibinfo{pages}{39}
  (\bibinfo{year}{2005}), \bibinfo{note}{quant-ph/0410199}.

\bibitem[{\citenamefont{Bacon}(2006)}]{Baco06}
\bibinfo{author}{\bibfnamefont{D.}~\bibnamefont{Bacon}},
  \bibinfo{journal}{Phys. Rev. A} \textbf{\bibinfo{volume}{73}},
  \bibinfo{pages}{012340} (\bibinfo{year}{2006}),
  \bibinfo{note}{quant-ph/0506023}.

\bibitem[{\citenamefont{Gottesman}(2013)}]{Gott13}
\bibinfo{author}{\bibfnamefont{D.}~\bibnamefont{Gottesman}},
  \bibinfo{journal}{arXiv:1310.2984}  (\bibinfo{year}{2013}).

\end{thebibliography}


\begin{thebibliography}{}
\expandafter\ifx\csname url\endcsname\relax
  \def\url#1{\texttt{#1}}\fi
\expandafter\ifx\csname urlprefix\endcsname\relax\def\urlprefix{URL }\fi
\providecommand{\bibinfo}[2]{#2}
\providecommand{\eprint}[2][]{\url{#2}}

\end{thebibliography}

\end{document}

\title{Supplementary Information for: ``Scalable in-situ qubit calibration during repetitive error detection''}

\author{J. Kelly}
\affiliation{Google Inc., Santa Barbara, California 93117, USA}

\author{R. Barends}
\affiliation{Google Inc., Santa Barbara, California 93117, USA}

\author{A. G. Fowler}
\affiliation{Google Inc., Santa Barbara, California 93117, USA}

\author{A. Megrant}
\affiliation{Google Inc., Santa Barbara, California 93117, USA}

\author{E. Jeffrey}
\affiliation{Google Inc., Santa Barbara, California 93117, USA}

\author{T. C. White}
\affiliation{Google Inc., Santa Barbara, California 93117, USA}

\author{D. Sank}
\affiliation{Google Inc., Santa Barbara, California 93117, USA}

\author{J. Y. Mutus}
\affiliation{Google Inc., Santa Barbara, California 93117, USA}

\author{B. Campbell}
\affiliation{Department of Physics, University of California, Santa
Barbara, CA 93106, USA}

\author{Yu Chen}
\affiliation{Google Inc., Santa Barbara, California 93117, USA}

\author{Z. Chen}
\affiliation{Department of Physics, University of California, Santa
Barbara, CA 93106, USA}

\author{B. Chiaro}
\affiliation{Department of Physics, University of California, Santa
Barbara, CA 93106, USA}

\author{A. Dunsworth}
\affiliation{Department of Physics, University of California, Santa
Barbara, CA 93106, USA}

\author{E. Lucero}
\affiliation{Google Inc., Santa Barbara, California 93117, USA}

\author{M. Neeley}
\affiliation{Google Inc., Santa Barbara, California 93117, USA}

\author{C. Neill}
\affiliation{Department of Physics, University of California, Santa
Barbara, CA 93106, USA}

\author{P. J. J. O'Malley}
\affiliation{Department of Physics, University of California, Santa
Barbara, CA 93106, USA}

\author{C. Quintana}
\affiliation{Department of Physics, University of California, Santa
Barbara, CA 93106, USA}

\author{P. Roushan}
\affiliation{Google Inc., Santa Barbara, California 93117, USA}

\author{A. Vainsencher}
\affiliation{Department of Physics, University of California, Santa
Barbara, CA 93106, USA}

\author{J. Wenner}
\affiliation{Department of Physics, University of California, Santa
Barbara, CA 93106, USA}

\author{John M. Martinis}
\affiliation{Google Inc., Santa Barbara, California 93117, USA}
\affiliation{Department of Physics, University of California, Santa
Barbara, CA 93106, USA}

\maketitle

\section{Determining patterns of independent parameter groups}
\subsection{Theoretical ideal case: Repetition code}

We consider the repetition code algorithm in the case where gates are non-ideal and may generate errors, but only consider the case where gate errors occur on the same qubit as the gate is performed. We model an imperfect Clifford gate as an ideal gate followed by some probability of an error X, Y or Z gate. Using this treatment, we determine where gate errors from each gate in the algorithm will be detected in Fig.~\ref{fig:pattern}a. Single qubit gate errors on measurement qubits are detected on the same qubit. Single qubit gate errors on data qubits are detected on neighboring measurement qubits. Given a CZ between a measurement qubit and a data qubit, errors from that CZ will be detected on both measurement qubits neighboring the data qubit. With these propagations in mind, we can partition the system into different patterns, with each pattern containing multiple groupings. Each grouping contains gates and qubits: errors from gates within the grouping will not propagate to qubits outside of that grouping. Each grouping always contains measurement qubits; their detection fraction $\zeta$ is used as a metric for gate performance. Then, each grouping can have its constituent gates optimized independently, and all groupings in a pattern can be optimized in parallel. 

\begin{figure}%[!]
    \centering
    \includegraphics[width=0.48\textwidth]{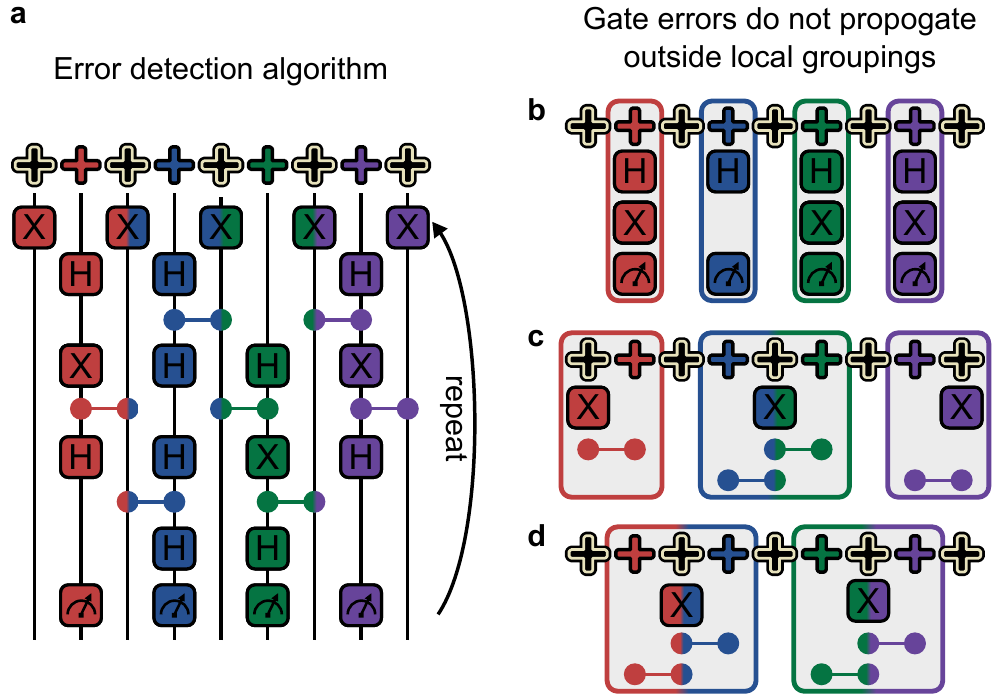}
    \caption{\textbf{Hardware patterns and local groupings.} (a) Error detection circuit, where gate color corresponds to which measurement qubit will detect errors from that gate. (b) First hardware pattern, where each of four groupings contains a measurement qubit and the single-qubit operations for that qubit. Errors from these gates will not propagate back to neighboring qubits. The relative $\zeta$ from the measurement qubit can be used to inform changes in gate parameters. (c, d) Second and third hardware pattern, where each grouping contains one data qubit and up to two measurement qubits. Groupings contain single-qubit gates on measurement qubits and CZ gates. The $\zeta$ of measurement qubits within each grouping can be used to inform gate parameters for gates within the grouping.}
    \label{fig:pattern}
\end{figure}

The first hardware pattern is shown in Fig~\ref{fig:pattern}b, where each grouping contains a measurement qubit and the single qubit operations for that qubit. The second hardware pattern, shown in Fig~\ref{fig:pattern}c, contains every other data qubit, and the neighboring measurement qubits. These groupings contain single qubit gates on the data qubits, and the CZ gates that involve that data qubit. The third hardware pattern shown in Fig~\ref{fig:pattern}d is the same, with the compliment set of data qubits. Between the three pattern and their groupings, we can access all gates on all qubits shown in Fig.~\ref{fig:pattern}a. 

To optimize all gates on all qubits, we would first (i) pick a hardware pattern, (ii) pick a gate to optimize within that hardware pattern and (iii) optimize that gate using ADEPT. We would then repeat i-iii with a different gate until we have covered all gates in all hardware patterns,

\subsection{Experimental case}

Any realistic system will deviate from the theoretically ideal case to some degree. For example, crosstalk may make gate parameters that should be local to one qubit effect the error rates of nearby qubits. However, this does not mean that ADEPT is not viable in such a system as long as crosstalk has a finite extent and is small enough. One can simply choose groups that contain qubits separated by the effective crosstalk distance. This will incur an additional overhead in the number of patterns used, but preserves $O(1)$ scaling with system size.

All non-idealities can be determined experimentally and algorithmically with a straightforward prescription. Assuming that a future large-scale system is composed of a cell of qubits with particular parameters (e.g. frequency, coupling) that is repeated throughout the computer, we can simply sweep each parameter of each gate of each qubit one at a time and record the corresponding change in error rate for the system. After going through all cases, and assuming that no parameter effects the entire system simultaneously, we will have a characteristic set of responses which can be used to determine independent patterns.

\section{Emulating continuous error detection}

Our current classical and quantum hardware is not yet suited for continuously running quantum error correction. In particular, our control software is not yet designed to update on-the-fly, and the qubits will likely require leakage reset to be able to operate with high fidelity for long numbers of rounds. However, we can still verify the underlying the principles behind ADEPT by emulating a continuously running system.

\begin{figure}%[!]
    \centering
    \includegraphics[width=0.48\textwidth]{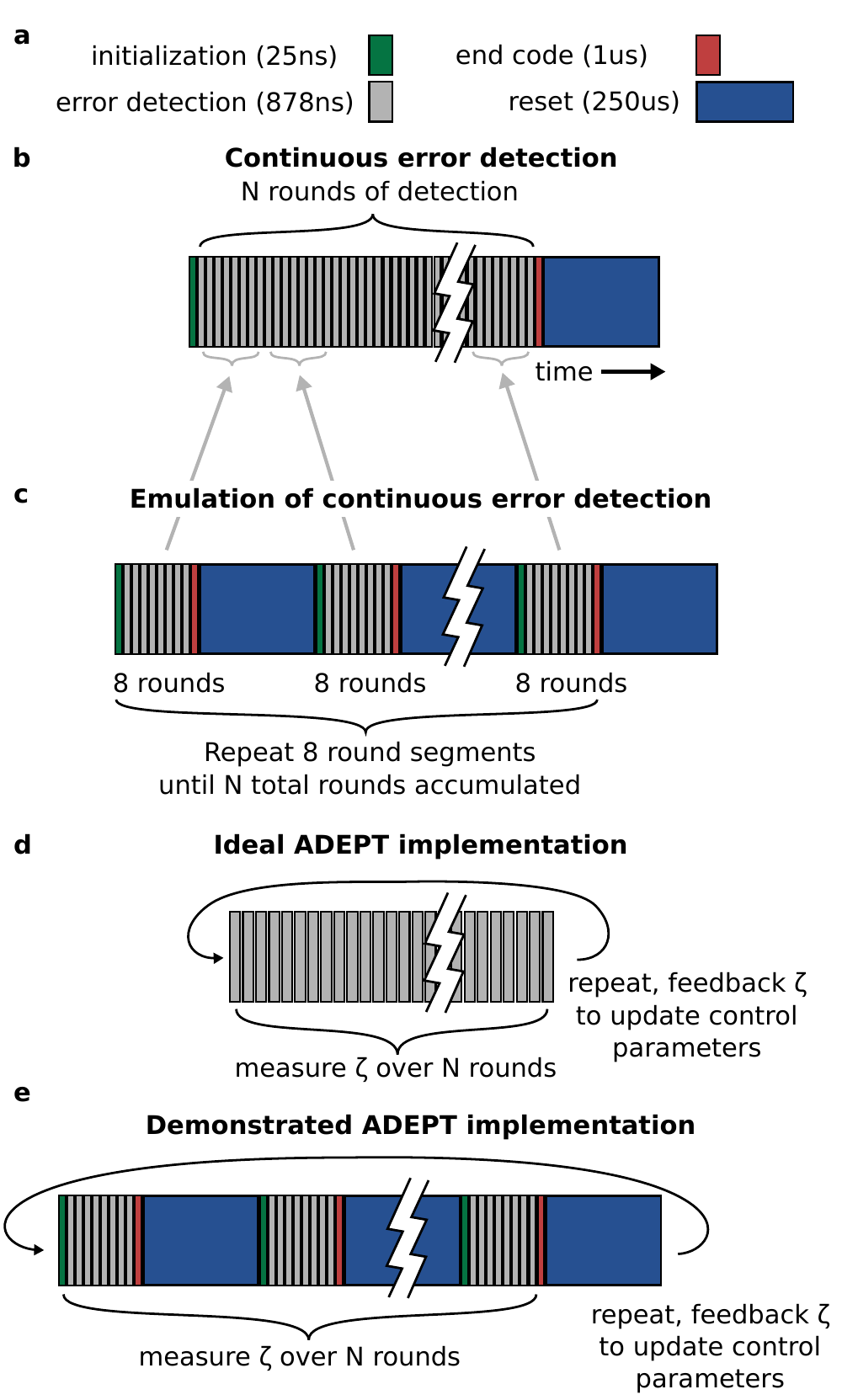}
    \caption{\textbf{Emulation of continuosuly running error detection.} (a) Error detection experiments are composed of four steps: initialization, error detection, ending the code, and system reset. (b) $N$ rounds of continuous error detection would involve initialization, $N$ rounds of error detection, then ending the code and resetting the system. (c) We emulate continuous error detection by running an ensemble of experiments that each contain eight rounds of detection, such that we accumulate $N$ total error detection rounds. (d) Ideally, ADEPT would run without ever needing to end the code of reset the system. Statistics over $N$ rounds of detection would be used to calculate $\zeta$, and then parameters would be updated on-the-fly. (e) We emulate this behavior by running ensembles of experiments to gather $N$ rounds of error detection to compute $\zeta$, and then updating parameters between ensembles. In main text Figures 3 and 4, the inserted bias is updated between ensembles of experiments.}
    \label{fig:emulation}
\end{figure}

To simulate $N$ rounds of continuously running error detection, we use an accumulation of experiments that each consist of eight rounds of error detection, see Fig.~\ref{fig:emulation}. We repeat many experiments, each with eight rounds of detections, until we have accumulated $N$ error detection cycles in total. 

It is of interest to estimate how fast a future continuously running quantum computer would be able to update parameters using ADEPT. We do this by quantifying ``emulation'' time; the timescale that a future computer will run at by subtracting out the time it takes to initialize, end and reset the code. For example, 36 ``emulated'' rounds of detection would take $\tau_{\textrm{emulated}} = 878 \textrm{ns} \cdot 36 = 31.6 \mu \textrm{s}$ on a future device, where our experiment would take $\tau_{\textrm{experiment}} = 878\textrm{ns} \cdot 36 + 3\cdot(25\textrm{ns} + 1
\mu \textrm{s} + 250\mu \textrm{s}) = 785 \mu \textrm{s}$.

\section{Bias-tracking algorithm}
\subsection{Quadratic error model}
In principle, a variety of algorithms could be used to track the bias drift of the qubit, but we use a simple algorithm based on a parabolic error model, as it is easy to understand analytically. Our model assumes

\begin{equation}
\zeta = a(x-x_0)^2 + \zeta_0
\label{eq:error_model}
\end{equation}

where $\zeta$ is the instantaneous detection rate, $\zeta_0$ is the background detection rate, $a$ relates the bias error to $\zeta$ (determined prior to the experiment), $x$ is the chosen bias parameter, and $x_0$ is the ideal bias value. Ideally, $x=x_0$. However, the ideal value can evolve in time from $x_0 \rightarrow x_1$, where $x_0$ becomes our most recent guess of the new ideal value $x_1$. We wish to determine $\delta x = x_1 - x_0$, the offset of our latest value to the ideal so that we can update our bias compensation.

\subsection{Where to sample}

We wish to determine $\delta_x$ while also keeping the instantaneous $\zeta$ to the base detection rate $\zeta_0$. In order to determine $\delta x$, we sample at points $x_0 \pm \Delta x$ and fit to a parabolic model. Importantly, sampling $x$ away from the optimum will increase $\zeta$, so there is a tradeoff in choosing $\Delta x$ to produce a large enough signal, and keeping the instantaneous $\zeta$ of the qubits compared to the base error rate. 

We aim to operate in the regime that $\Delta x \gg \delta x$, so let us set $\delta x = 0$ temporarily. suppose we choose to tolerate a fractional increase F in the base $\zeta_0$. We wish to determine $\Delta x$, how far we should sample from our most recent optimum value $x_0$ while only incurring an additional detection of $F \zeta_0$,

\begin{align}
\zeta &= \zeta_0(1+F) = a (\Delta x)^2 + \zeta_0 \\
\Delta x &= \sqrt{\zeta_0F/a}.
\label{eq:sample}
\end{align}

\subsection{Updating the bias}

Once we have sampled at $x= x_0\pm \Delta$ giving us $\zeta_{\pm}$, we can determine $\delta x$.

\begin{align}
\zeta_+& =a(\Delta x + x_0 - x_1)^2 + \zeta_0 \\
 \zeta_-&=a(-\Delta x + x_0 - x_1)^2 + \zeta_0 \\
\delta x &= \frac{\zeta_- - \zeta_+}{4a\Delta x}
\label{eq:deltax}
\end{align}

\begin{figure}[t!]
    \centering
    \includegraphics[width=0.48\textwidth]{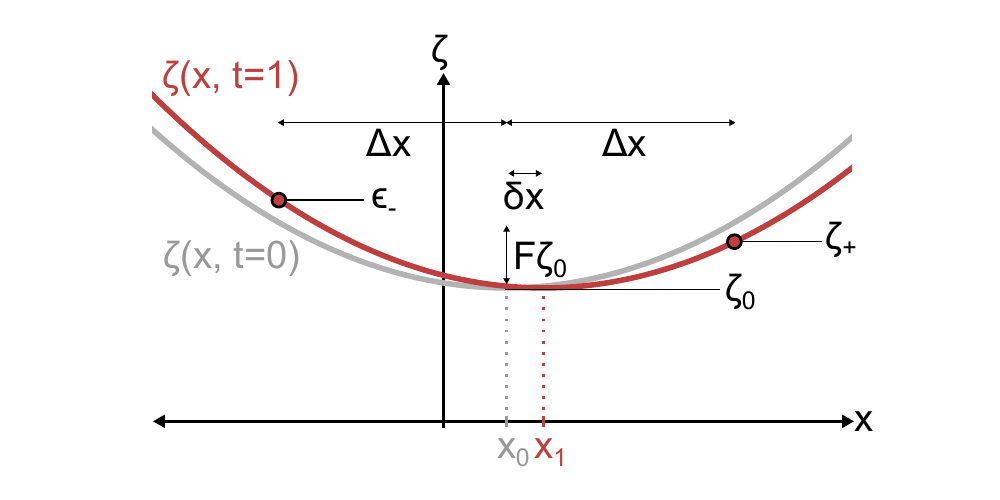}
    \caption{\textbf{Parabolic error model.} Errors take the form of Eq.~\ref{eq:error_model}. After some time, the ideal parameter $x_0$ at $t=0$ drifts to $x_1$ at $t=1$. Detection fraction $\zeta$ is sampled at $x_0 \pm \Delta x$ defined in Eq.~\ref{eq:sample}. $\zeta$ is increased by $F \zeta_0$ when using this sampling. After fitting to a parabolic error model, a new parameter $x_1$ is determined by adding $\delta x$ as defined in Eq~\ref{eq:deltax}.}
    \label{fig:pattern}
\end{figure}

\subsection{Sampling statistics}

As we sample for a finite time, we make imperfect measurements of $\zeta_{\pm}$. These will translate into noise in our determined $\delta x$ parameter that we use to update the bias. We want to operate in the regime where for $\delta x$ its standard error $SE_{\delta x}$ is much less than $\Delta x$, as we expect $\delta x \ll \Delta x$ for slow drift. Our goal is to accumulate enough statistics from sampling $N$ times to achieve the condition for small relative noise $P \ll 1$.

\begin{align}
P &= \frac{SE_{\delta x}}{\Delta x}
\label{eqn:se}
\end{align}

Given that $\zeta_{\pm}$ will be sampling the binomial distribution of detection fraction $\zeta$, the standard error of the mean for $\zeta_{\pm}$ is  

\begin{align}
SE_{\zeta_{\pm}}=\frac{\sqrt{\zeta_{\pm}(1-\zeta_{\pm})}}{\sqrt{N}}
\label{eq:se}
\end{align}

where N is the number of experiments. Adding the standard deviations in quadrature for $\zeta_{\pm}$ and using Eq.~\ref{eq:deltax} we find

\begin{align}
SE_{\delta x}=\frac{\sqrt{2\zeta_{\pm}(1-\zeta_{\pm})}}{\sqrt{N}4 a \Delta x}
\label{eq:seeps}
\end{align}

Solving Eq.~\ref{eqn:se} and Eq.~\ref{eq:seeps} using Eq.~\ref{eq:sample} and $\zeta_{\pm}=\zeta_0(1+F)$ one finds the condition for N

\begin{align}
N \approx \frac{1}{8 P^2 F^2 \zeta_0}
\end{align}

Plugging in the relevant parameters $\zeta_0=0.15, F=0.1$ and taking $P=25$ we get $N\approx50,000$ where as we used $N=48,000$ for the experiment in main text Figure~4.

\subsection{Sampling speed}

Assuming the 1.1~MHz error detection rate, $N=48,000$ measurements for each $\zeta$ measurement, two measurements per update, and three qubit patterns to cycle between as in main text Figure 4, frequency drift as fast as 0.3~Hz could be compensated for every qubit in a continuously running repetition code experiment.

\bibliographystyle{naturemag}
\bibliography{bibliography}